\documentstyle[12pt]{article}
\newcommand{\be}{\begin{equation}}
\newcommand{\ee}{\end{equation}}
\newcommand{\bea}{\begin{eqnarray}}
\newcommand{\eea}{\end{eqnarray}}
\newcommand{\CR}{\nonumber \\}
\newcommand{\A}{\alpha}
\newcommand{\B}{\beta}

\newcommand{\LM}{\Lambda}
\newcommand{\Om}{\Omega}
\newcommand{\lm}{\lambda}
\newcommand{\pa}{\partial}
\newcommand{\cL}{{\cal L}}
\newcommand{\cF}{{\cal F}}
\newcommand{\DE}{\Delta}
\newcommand{\vt}{\vartheta}
\newcommand{\tcL}{\widetilde{\cal L}}
\newcommand{\tOm}{\widetilde{\Omega}}
\newcommand{\tb}{\tilde{b}}
\begin{document}
\baselineskip=0.7cm
\renewcommand{\thefootnote}{\fnsymbol{footnote}}
\begin{titlepage}
\begin{flushright}
hep-th/9703180 \\
UTHEP-359 \\
March, 1997
\end{flushright}
\bigskip
\bigskip
\begin{center}
{\Large \bf Picard-Fuchs Equations and Prepotential \\ in $N=2$ 
Supersymmetric $G_{2}$ Yang-Mills Theory}

\bigskip

Katsushi Ito 
\medskip

{\it Institute of Physics, University of Tsukuba\\ Ibaraki 305, Japan}
\end{center}

\bigskip

\bigskip

\begin{abstract}
We study the low-energy effective theory of $N=2$ supersymmetric Yang-Mills 
theory with the exceptional gauge group $G_{2}$.
We obtain the Picard-Fuchs equations for the $G_{2}$ spectral curve and 
compute multi-instanton contribution to the prepotential.
We find that the spectral curve is consistent with the microscopic 
supersymmetric instanton calculus.
It is also shown that $G_{2}$ hyperelliptic curve does not reproduce
the microscopic result.

\end{abstract}
\end{titlepage}
\renewcommand{\thefootnote}{\arabic{footnote}}
\setcounter{footnote}{0}

Seiberg and Witten found the exact solution for the low energy effective 
theory of $N=2$  supersymmetric $SU(2)$ Yang-Mills theory\cite{SeWi}.
Their $SU(2)$ solution has been generalized to other gauge groups
\cite{KlLeThYa}-\cite{MaWa}.
The low-energy effective theory in the Coulomb branch 
of $N=2$ supersymmetric Yang-Mills theory
with gauge group $G$ of rank $r$
is described by $r$  $N=1$ $U(1)$ vector multiplets 
$W_{\A}^{i}=(\lm_{\A}^{i},v_{m}^{i})$ and $r$ $N=1$ hypermultiplets 
$A^{i}=(a^{i},\psi_{\A}^{i})$ ($i=1,\ldots,r$).
The low energy effective action is determined by a single holomorphic 
function $\cF(a)$, called the prepotential. 
In the semi-classical region, the prepotential $\cF(a)$ is 
expressed as
\be
\cF(a)={i\over 4\pi} \sum_{\A\in\DE_{+}} (\A,a)^{2}
\log {(\A,a)^{2}\over \LM^{2}}+{1\over2}\tau_{0} \sum_{i=1}^{r}a^{i}a^{i}
+\sum_{n=1}^{\infty} \cF_{n}(a) \LM^{2 h^{\vee}n}.
\ee
where $\Delta_{+}$ denotes the set of positive roots of the Lie algebra 
of $G$, $h^{\vee}$ the dual Coxeter number and 
$\LM$ the dynamically generated mass scale.
The coefficient $\cF_{n}(a)$ comes from the $n$-instanton 
contribution.

In the exact solution, the Higgs fields $a^{i}$ and their duals 
$a_{D i}=\pa \cF(a)/ \pa a^{i}$ are represented by the contour integral
of the meromorphic one-form (the Seiberg-Witten differential) 
$\lm_{SW}$ on certain algebraic curve:
\be
a^{i}=\int_{A_{i}}\lm_{SW}, \quad a_{D i}=\int_{B_{i}}\lm_{SW},
\ee
where $A_{i}$ and $B_{i}$ are 1-cycle on the algebraic curve.
For classical gauge groups, it is known that the hyperelliptic 
curve provides the  exact solution which satisfies several consistency 
conditions\cite{KlLeThYa}.
This type of curves has been generalized to exceptional type gauge groups
by embedding them to certain classical gauge groups 
\cite{DaSu2}-\cite{ABOALI}.

Generalizing the work by Gorskii et. al. \cite{Gor},
Martinec and Warner \cite{MaWa} constructed the algebraic curves for 
any simple gauge group from the spectral
curve of the periodic Toda lattice associated with the dual affine Lie algebra.
For classical gauge groups, the spectral curve is shown to agree with
the hyperelliptic type. 
For exceptional type gauge groups, however, the spectral curve has different 
from the hyperelliptic curves and shows different strong coupling physics.
Since the singularity structure in the strong coupling region 
determines instanton terms in the prepotential by analytic continuation, 
the calculation of $n$-instanton contributions provides a non-trivial 
test to the exact solutions. 

Recently, Sasakura and the present author  \cite{ItSa} calculated the 
one-instanton effect $\cF_{1}(a)$ for any simple Lie group by 
using the microscopic supersymmetric instanton calculus\cite{Inst}.
These effects have been shown to agree with the exact solutions in the
case of classical gauge groups \cite{DhKrPh,MaSu}. 

The purpose of the present paper is to study the exact solution for 
exceptional type gauge groups.
We will consider the $G_{2}$ type Lie group as the simplest example. 
We study the exact solution by investigating the Picard-Fuchs equations
that the period $\oint \lm_{SW}$ obeys. 
The Picard-Fuchs equation has been extensively studied for classical 
gauge groups \cite{Cer}-\cite{EwFoTh}. 
By solving these differential equations in the semi-classical region, we 
obtain instanton correction to the prepotential explicitly.

Another non-trivial consistency check for $G_{2}$ gauge group 
has been proposed by Landsteiner et al. \cite{LaPiGi}. 
They apply the method of confining phase superpotential \cite{ElFoGiInRa}
 to $G_{2}$ 
and find that the discriminant of the spectral curve is consistent 
with that from the superpotential.
The present approach provides another nontrivial and quantitative
check to the exact solutions.

The $G_{2}$ type Lie algebra ($h^{\vee}=4$) 
 contains six positive roots.
Let $\A_{1}=(\sqrt{2},0)$ and $\A_{2}=(-{1\over \sqrt{2}},{1\over \sqrt{6}})$
be simple roots.
Among the the positive roots, $\A_{1}$, $\A_{1}+3\A_{2}$ and $2\A_{1}+3\A_{3}$
are long roots with squared length 2.
Remaining roots $\A_{2}$, $\A_{2}+\A_{3}$ and $\A_{1}+2\A_{2}$ are 
short roots with squared length 2/3.
The fundamental weights $\lm_{1}$ and $\lm_{2}$ are defined by
$\lm_{1}=2\A_{1}+3\A_{2}$ and $\lm_{2}=\A_{1}+2\A_{2}$.

The representation with the highest weight $\lm_{2}$ is seven-dimensional
and may be embedded into that of the vector representation of the 
Lie algebra $so(7)$.
One may use this embedding to construct the hyperelliptic curve 
for the $G_{2}$ gauge group from the gauge group $SO(7)$ with $N_{f}=1$ flavor
\cite{DaSu2,AlArMa}:
\be
y^{2}=((x^2-\tb_{1}^2)(x^2-\tb_{2}^2)(x^2-\tb_{3}^2))^2 -\LM^{8} x^{4},
\label{eq:hyp}
\ee
where
\bea
\tb_{1}&=& b_{2}, \CR
\tb_{2}&=& b_{1}-b_{2}, \CR
\tb_{3}&=& -b_{1}+2 b_{2},
\eea
and $b_{i}=(\lm_{i},a)$ ($i=1,2$).
This curve are shown to have correct monodromy in the weak coupling region.
One can calculate one-instanton contribution to the prepotential by
the restriction of $SO(7)$ $N_{f}=1$ theory \cite{DhKrPh} to $G_{2}$.
The result reads
\be
\cF_{1}(b)\LM^{8}
= -{i \LM^{8}\over 2^6\pi} { \tb_{1}^2 (\tb_{2}^2-\tb_{3}^2)^2+
           \tb_{2}^2 (\tb_{1}^2-\tb_{3}^2)^2+
           \tb_{3}^2 (\tb_{1}^2-\tb_{2}^2)^2\over
 (\tb_{1}^2-\tb_{2}^2)^2 (\tb_{2}^2-\tb_{3}^2)^2 (\tb_{1}^2-\tb_{3}^2)^2}.
\ee
This is shown to be
\be
\cF_{1}(b)\LM^{8}=-{\LM^{8}i\over 2^6\pi}
 \left(   {{9\over2}\over (2b_{1}-3b_{2})^2 (b_{1}-3 b_{2})^2 b_{1}^2}
        +{{1\over2}\over (b_{1}-2b_{2})^2 (b_{1}- b_{2})^2 b_{2}^2}
\right) .
\ee
The first term is made of the contribution from the long roots.
The second term contains the short roots only.
One cannot expect the latter type of singularity from the microscopic 
instanton calculation.
In fact, the microscopic instanton calculus \cite{ItSa} 
shows that the one-intanton 
contribution to the prepotential is given by
\be
\cF_{1}^{inst.}\LM_{PV}^{8}=-{i\LM_{PV}^{8}\over 2^{4}\pi}
 {9 \over (2b_{1}-3b_{2})^2 (b_{1}-3 b_{2})^2 b_{1}^2},
\label{eq:micro}
\ee
where $\LM_{PV}$ is the scale parameter defined in the Pauli-Villars
regularization scheme.
Therefore the hyperelliptic curve (\ref{eq:hyp}) predicts additional
singularities arising from the zero vacuum expectation value of a
short root.
The one-instanton term does not coincide with the result from the 
microscopic instanton calculation.

We next study the exact solution associated with the spectral curve which 
comes from the $(G^{(1)}_{2})^{\vee}$ Toda lattice\cite{MaWa}.
The spectral curve for $G_{2}$ reads
\be
3\left(z-{\mu\over z}\right)^2
-x^8+2  u x^6-
\left[ u^2 +\left(z+{\mu\over z}\right)\right] x^4
+\left[ v+ 2 u \left(z+{\mu\over z}\right) \right] x^2=0.
\label{eq:spec}
\ee
The Seiberg-Witten differential is given by
\be
\lm_{SW}=x {d z\over z}.
\ee
It is convenient to introduce a new variable $y=z+{\mu\over z}$. Then the
Seiberg-Witten one-form take the form
\be
\lm_{SW}=x {d y\over \sqrt{y^2-4\mu}}.
\ee
Here $y$ satisfies the quadratic equation
\be
3 y^{2}-c_{1} y-c_{2}=0,
\label{eq:quad}
\ee
where
\bea
c_{1}&=& 6 x^{4}-2 u x^{2}, \\
c_{2}&=& x^{8} -2 u x^{6}+u^{2} x^{4}-v x^{2} + 12 \mu .
\eea
The equation (\ref{eq:quad}) have two solutions:
\be
y={1\over 6} (c_{1}\pm \sqrt{c_{1}^2+12 c_{2}}).
\ee
In the following analysis we take plus sign without loss of generality.
The canonical holomorphic one-forms on the spectral curve are given by 
taking derivative of $\lm_{SW}$ with respect to $u$ and $v$:
\be
{\pa\lm_{SW}\over \pa t}=-{{\pa y\over \pa t}\over \sqrt{y^2-4\mu}}dx
-{\pa\over \pa x}\left(
{x {\pa y\over \pa t} \over \sqrt{y^2-4\mu}}
\right) dx,
\ee
where $t=u$ or $v$.

We look for the Picard-Fuchs equations of the form
\be
a_{tt}{\pa^{2}\lm_{SW}\over \pa t^2}
+a_{uv} {\pa^{2}\lm_{SW}\over \pa u\pa v}
+a_{u} {\pa\lm_{SW}\over \pa u}
+a_{v} {\pa\lm_{SW}\over \pa v}
-\lm_{SW}=d\left( { f+g \sqrt{c_{1}^2+12 c_{2}} \over 
\sqrt{c_{1}^2+12 c_{2}} \sqrt{y^2-4\mu}} \right)
\ee
where $t=u$ or $v$. $f$ and $g$ are polynomials of fourth order in $x$.
After some computations we find that the differential equations
for the periods $\Pi=\oint \lm_{SW}$ are given by $\cL_{i}\Pi=0$ 
($i=1,2$) where
\bea
\cL_{1}&=& ({8\over 3} u^3 v-36 v^2+960 u^2\mu) \pa_{v}^2
           +({8\over 3} u^4 -24 u v+2304 \mu) \pa_{u}\pa_{v}
           +(4 u^3-24 v)\pa_{v}-1,
\CR
\cL_{2}&=& {2 (720 u^2\mu+2 u^3 v-27 v^2)\over -v u+24 \mu}  \pa_{u}^2
  +{4 (256 u^4 \mu-3 u^2 v^2-720 v u \mu+13824 \mu^2)\over -v u+24 \mu}
 \pa_{u}\pa_{v} \CR
& & -{6 (-256 u^3\mu+96\mu v+5 v^2 u)\over  -v u+24 \mu} \pa_{v}-1.
\eea
Let us define differential operators $\tcL_{i}$ by
\bea
\tcL_{1}&=& (1-2 u^2 )\cL_{1}-2 u^2 \cL_{2},
\CR
\tcL_{2}&=& {v\over u} (\cL_{1}-\cL_{2}).
\eea
These operators are convenient for studying solutions in the 
semi-classical region.
In fact, these differential equations are written in the 
form of hypergeometric system by introducing new variables
$x={v\over u^{3}}$ and $y={\mu u^2\over v^2}$:
\bea
\tcL_{1}&=& 1024 y (\vt_{x}-2\vt_{y})(\vt_{x}-2\vt_{y}-1)
+2304 x y (-3\vt_{x}+2\vt_{y}) (\vt_{x}-2\vt_{y})
-(8\vt_{y}+1)^2,
\CR
\tcL_{2}&=&-32 x y (\vt_{x}-2\vt_{y})(\vt_{x}-2\vt_{y}-1)
+2 x (-3\vt_{x}+2\vt_{y}) (-3\vt_{x}+2\vt_{y}-1) \CR
& & +{2\over 3} (\vt_{x}-2\vt_{y}) (-4\vt_{x}+1),
\label{eq:pf2}
\eea
where $\vt_{x}=x\pa_{x}$ and $\vt_{y}=y\pa_{y}$ are the Euler derivatives.

Now we construct the solution of the Picard-Fuchs equations 
$\tcL_{i}\Pi=0$ ($i=1,2$) in the semi-classical region where $\mu$ is small. 
Consider a formal power series solution around $(x,y)=(0,0)$ of the 
form
\be
\omega_{\A,\B}(x,y)=\sum_{m,n\geq 0} c_{\A,\B}(m,n) x^{m+\A} y^{n+\B},
\ee
where $c_{\A,\B}(0,0)=1$.
The indicial equations become
\bea
(8\B+1)^2&=& 0, \CR
(\A-2\B)(-4\A+1)&=&0.
\label{eq:ind}
\eea
The equations (\ref{eq:ind}) have two degenerate solutions 
$(\A,\B)=(-1/4,-1/8)$ and $(1/4,-1/8)$.
Applying the Frobenius method, we find two other solutions of 
logarithmic type.
Finally four solutions of the the Picard-Fuchs equations (\ref{eq:pf2})
are given by
\bea
\Om_{1}(x,y)&=& \omega_{-1/4,-1/8}(x,y), \CR
\Om_{2}(x,y)&=& \omega_{1/4,-1/8}(x,y), \CR
\Om_{D1}(x,y)&=&
\left. ({\pa\over\pa\A}+{1\over2}{\pa\over\pa\B})\omega_{\A,\B}(x,y)
\right|_{(\A,\B)=(-1/4,-1/8)}  \CR
&=& \Om_{1}(x,y)\log (x y^{1/2})
+\sum_{m,n\geq 0} \left.({\pa\over\pa\A}+{1\over2}{\pa\over\pa\B})
c_{\A,\B}(m,n)\right|_{(\A,\B)=(-1/4,-1/8)} x^{m+\A} y^{n+\B}, \CR
\Om_{D2}(x,y)&=&
\left. {\pa\over\pa\B}\omega_{\A,\B}(x,y)\right|_{(\A,\B)=(1/4,-1/8)} \CR
&=& \Om_{2}(x,y)\log (y)
+\sum_{m,n\geq 0} \left.{\pa\over\pa\B}
c_{\A,\B}(m,n)\right|_{(\A,\B)=(1/4,-1/8)} x^{m+\A} y^{n+\B}, \CR
\eea
Here the coefficients $c_{\A,\B}(m,n)$ obey the recursion relations
\bea
c_{\A,\B}(m,n)&=& A_{\A,\B}(m,n) c_{\A,\B}(m,n-1)
                  +B_{\A,\B}(m,n) c_{\A,\B}(m-1,n-1), \CR
c_{\A,\B}(m,n)&=&  C_{\A,\B}(m,n) c_{\A,\B}(m-1,n)
                  +D_{\A,\B}(m,n) c_{\A,\B}(m-1,n-1),
\eea
where
\bea
A_{\A,\B}(m,n)&=& 
{1024 (m-2 n+2 +\A-2\B) (m-2 n+1 +\A-2\B)\over (8 n+1+\B)^2}, \CR
B_{\A,\B}(m,n)&=& {2304 (-3 m+2 n +1-3\A+2\B) (m-2 n +1 +\A-2\B)
                   \over (8 n+1+\B)^2}, \CR
C_{\A,\B}(m,n)&=& {-3 (-3 m+2 n +3-3\A+2\B) (-3 m+2 n +2-3\A+2\B)
                   \over (m-2n +\A-2\B) (-4n+1-4\A)}, \CR
D_{\A,\B}(m,n)&=& { 48 (m-2 n +1+\A-2\B)(m-2 n +\A-2\B)
                   \over (m-2n +\A-2\B) (-4n+1-4\A)}. \CR
\eea
Therefore the coefficients are obtained recursively
\bea
c_{\A,\B}(m,0)&=& \left({27\over 4}\right)^{m}
                  {(\A-{2\over 3}\B)_{m} (\A-{2\over3}\B+{1\over3})_{m}
                   \over (\A-2\B+1)_{m} (\A+{3\over4})_{m}}, \CR
c_{\A,\B}(m,1)&=& A_{\A,\B}(m,1) c_{\A,\B}(m,0)
                  +B_{\A,\B}(m,1) c_{\A,\B}(m-1,0), \CR
c_{\A,\B}(m,2)&=& A_{\A,\B}(m,2) c_{\A,\B}(m,1)
                  +B_{\A,\B}(m,2) c_{\A,\B}(m-1,1),
\eea
etc. where $(a)_{m}=\Gamma(a+m)/\Gamma(a)$.
The first few terms of the series expansions of the solutions are given by
\bea
\tOm_{1}(x,y)&=& \sqrt{u}-{3\over8} {v\over u^{5/2}}
-{105\over 128} {v^{2}\over u^{11/2}}
+{15\over2}{\mu\over u^{7/2}}
 +{693\over 16} {\mu v\over u^{13/2}}+\cdots, \CR
\tOm_{2}(x,y)&=& {\sqrt{u}\over v}+{v^{3/2}\over u^{4}}
+3  {v^{3/2}\over u^{4}}-4 {\mu u\over v^{3/2}}
-6{\mu \over u^{2}\sqrt{v}}+\cdots ,\CR
\tOm_{D1}(x,y)&=& \tOm_{1}(x,y)
\log {\mu^{1/2}\over u^{2}}+
{3\over 4} {v\over u^{5/2}}+{17\over 16} {v^{2}\over u^{11/2}}
-14{\mu\over u^{7/2}}-{2307\over 40} {\mu v\over u^{13/2}}+\cdots,
\CR
\tOm_{D2}(x,y)&=& \tOm_{2}(x,y)\log {\mu u^{2}\over v^{2}}
  -{5\over 3} {v^{3/2}\over u^{4}}-{53\over 10} {v^{5/2}\over u^{7}}
+8 {\mu u\over v^{3/2}}+36 {\mu \over u^{2}\sqrt{v}}+\cdots ,
\eea
where $\tOm_{i}=\mu^{1/8}\Om_{i}$ and $\tOm_{D i}=\mu^{1/8}\Om_{D i}$ 
($i=1,2$).
One may construct the classical solutions
\bea
b_{1}&=& \sqrt{3}\tOm_{1}-{\sqrt{3}\over 2}\tOm_{2}, \CR
b_{2}&=& {2\over \sqrt{3}} \tOm_{1}, \CR
b_{D1}&=& {i\over 2\pi} \sqrt{3}\tOm_{D2}+t_{0} (2 b_{1}-3 b_{2}), \CR
b_{D2}&=& {i\over 2\pi} \left( -2\sqrt{3}\tOm_{D1}
-{3\sqrt{3}\over2}\tOm_{D2}\right)+t_{0} (-3 b_{1}+6 b_{2}) ,
\eea
where $t_{0}$ is a constant which is obtained by evaluation of the 
contour integral.
But the value of $t_{0}$ is not necessary for the determination of the
instanton effects to the prepotential.
{}From these solutions, we may obtain the identities
\be
\sum_{i=1}^{2}(\pa_{t}b_{D i} b_{i}-b_{D i}\pa_{t}b_{i})
={i\over 4\pi}\delta_{t,u}
\label{eq:scale1}
\ee
where $t=u$ or $v$. 
Due to the complicated structure of poles in $\lm_{SW}$, it is difficult to
prove (\ref{eq:scale1}) directly in a similar way as \cite{So}.
We have explicitly checked (\ref{eq:scale1}) up to order $\mu^{5}$.
Hence the present results are exact up to 5-instanton level.
By integration  the identities (\ref{eq:scale1}) over $u$ or $v$,
we get the scaling equation\cite{Ma,So}
\be
{i u\over 4\pi}=\sum_{i=1}^{2} b_{i} {\pa \cF(b)\over \pa b_{i}}-2\cF(b).
\ee
This identity allows us to calculate the $n$-instanton effects 
explicitly.
The first three terms are given as
\bea
\cF_{1}(b)\LM^{8} &=& -{3^4 i \LM^{8}\over \pi} 
{1\over (2b_{1}-3b_{2})^2 (b_{1}-3 b_{2})^2 b_{1}^2}, \CR
\cF_{2}(b)\LM^{16}&=& -{3^{10} 5 i \LM^{16}\over 2\pi} 
{(b_{1}^2-3b_{1}b_{2}+3 b_{2}^2)^2
\over (2b_{1}-3b_{2})^6 (b_{1}-3 b_{2})^6 b_{1}^6}, \CR
\cF_{3}(b)\LM^{24}&=& -{2^3 3^{17}i\LM^{24}\over \pi} 
{(b_{1}^2-3b_{1}b_{2}+3 b_{2}^2)^4
\over (2b_{1}-3b_{2})^{10} (b_{1}-3 b_{2})^{10} b_{1}^{10}}
+{i 116\cdot 3^{12} \LM^{24}\over \pi}
{(b_{1}^2-3b_{1}b_{2}+3 b_{2}^2)
\over (2b_{1}-3b_{2})^{8} (b_{1}-3 b_{2})^{8} b_{1}^{8}} \CR
& & -{i 2^6 3^{6} \LM^{14}\over \pi}
{(b_{1}^2-3b_{1}b_{2}+3 b_{2}^2)
\over (2b_{1}-3b_{2})^{6} (b_{1}-3 b_{2})^{6} b_{1}^{6}
b_{2}^2 (b_{1}-b_{2})^2 (-b_{1}+2 b_{2})^2},
\label{eq:specinst}
\eea
where we put $\LM=\mu^{1/8}$.
If the parameter $\mu$ satisfies the relation
\be
\LM_{PV}^{8}=3^{2} 2^{4} \mu,
\ee
we find the one-instanton term in eqs. (\ref{eq:specinst}) agrees with
that from the microscopic calculation (\ref{eq:micro}).
As another consistency check, let us consider the  $SU(2)$ limit 
$b_{2}\rightarrow \infty$ with the matching condition
\be
\mu={b_{2}^{4} \LM_{SU(2)}^4\over 4},
\ee
and finite $\LM_{SU(2)}$.
In this limit, we can show that the above instanton series reduces to that of 
the prepotential for $SU(2)$ gauge group\cite{KLT,ItYa,Ma}.

Note that in the classical limit $\mu=0$, $\pm\tb_{i}$ ($i=1,2,3$) obey
the equation
\be
x^{6}-2 u x^4+u^2 x^2 -(-v+{4\over 27}u^3)=0,
\ee
which is obtained from the classical characteristic polynomial
$x^{6}-2 u x^4+u^2 x^2 -v$ by the transformation
\be
\left\{
\begin{array}{cl}
u&\rightarrow u, \\
v&\rightarrow -v+{4\over 27} u^3.
\end{array}
\right.
\ee
The necessity of this replacement of variables has been also noticed 
in ref. \cite{LaPiGi}.

In this paper we have studied the exact solutions represented by the
hyperelliptic and spectral curves for the exceptional gauge group $G_{2}$.
We have shown that the spectral curve  (\ref{eq:spec}) gives the 
prepotential which is consistent with the microscopic instanton
calculation.
But the hyperelliptic curve (\ref{eq:hyp}) does not agree with 
the microscopic result.
The present analysis suggests that the spectral curves provides
a systematic approach to the exact solutions to the 
Seiberg-Witten theory.
It is interesting to generalize the present analysis to 
other exceptional gauge groups. 
In particular,  $E_{6}$ type gauge groups would be treated in a similar way.
$E_{r}$ type gauge groups are particularly interesting in viewpoint of 
string duality, since the ALE fibration \cite{KlLeMaVaWa}
gives systematic construction of the spectral curve\cite{LeWa}.
The microscopic one-instanton calculation would provide quantitative 
test to the exact solutions and string duality in these cases.

\section*{Acknowledgments}
The author would like to thank N.~Sasakura for valuable discussions.
This work is supported in part by
the Grant-in-Aid for Scientific Research from
the Ministry of Education (No.~08740188 and No.~08211209).


\newpage
\begin{thebibliography}{99}
\bibitem{SeWi} N.~Seiberg and E.~Witten, Nucl. Phys. {\bf B426} (1994) 19;
ibid. {\bf B431} (1994) 484.

\bibitem{KlLeThYa}
A. Klemm, W. Lerche, S. Theisen and S. Yankielowicz,
Phys. Lett. {\bf B344} (1995) 169; \\
P.C. Argyres and A.E. Faraggi, Phys. Rev. Lett. {\bf 74} (1995) 3931; \\
U.H.~Danielsson and B.~Sundborg,   Phys. Lett. {\bf B358} (1995) 273; \\
A.~Brandhuber and K.~Landsteiner,   Phys. Lett. {\bf B358} (1995) 73; \\
A.~Hanany and Y.~Oz,   Nucl. Phys. {\bf B452} (1995) 283; \\
P.C.~Argyres, M.R.~Plesser and A.D.~Shapere,
Phys. Rev. Lett. {\bf 75} (1995) 1699; \\
J.A.~Minahan and D. Nemeschansky, Nucl. Phys. {\bf B464} (1996) 3; \\
P.C.~Argyres and A.D.~Shapere, Nucl. Phys. {\bf B461} (1996) 437; \\
A.~Hanany, Nucl. Phys. {\bf B466} (1996) 85.

\bibitem{DaSu2}
U.H.~Danielsson and B.~Sundborg, Phys. Lett. {\bf B370} (1996) 83.

\bibitem{AlArMa}
M. Alishahiha, F. Ardalan and F. Mansouri,
Phys. Lett. {\bf B381} (1996) 446.

\bibitem{ABOALI}
M.R.~Abolhasani, M.~Alishahiha and A.M.~Ghezelbash,
Nucl. Phys. {\bf B480} (1996) 279.

\bibitem{Gor}
A. Gorskii, I. Krichever, A. Marshakov, A. Mironov and A. Morozov,
Phys. Lett. {\bf B355} (1995) 466.

\bibitem{MaWa}
E.~Martinec and N.P.~Warner, Nucl. Phys. {\bf B459} (1996) 97.

\bibitem{ItSa} K.~Ito and N.~Sasakura, Phys. Lett. {\bf B382} (1996) 95;
Nucl. Phys. {\bf B484} (1997) 141.

\bibitem{Inst}
I.~Affleck, M.~Dine and N.~Seiberg, Nucl. Phys.  {\bf B241} (1984) 493; \\
V.A.~Novikov, M.A.~Shifman, A.I.~Vainshtein and V.I.~Zakharov,
Nucl. Phys.  {\bf B260} (1985) 157; \\
D.~Amati, K.~Konishi, Y.~Meurice, G.C.~Rossi and G.~Veneziano,
Phys. Rep. {\bf 162} (1988) 169.


\bibitem{DhKrPh}
E.~D'~Hoker, I.M.~Krichever and D.H.~Phong, hep-th/9609041; hep-th/9609145.

\bibitem{MaSu}
T.~Masuda and H.~Suzuki, hep-th/9609065.

\bibitem{LaPiGi}
K.~Landsteiner, J.M.~Pierre and S.B.~Giddings, 
Phys. Rev. {\bf D55} (1997) 2367.

\bibitem{ElFoGiInRa}
S.~Elitzur, A.~Forge, A.~Giveon, K.~Intriligator and E.~Rabinovici,
Phys. Lett. {\bf B379} (1996) 121; \\
S.~Terashima and S.-K.Yang, Phys. Lett. {\bf B391} (1997) 107; \\
T.~Kitao, hep-th/9611097; \\
T.Kitao, S.~Terashima and S.-K.Yang, hep-th/9701009; \\
A.~Giveon, O.~Pelc  and E.~Ravinovici, hep-th/9701045.
\bibitem{Cer}
A. Ceresole, R. D'Auria and S. Ferrara, Phys. Lett. {\bf B339}
(1994) 71.

\bibitem{KLT}
A. Klemm, W. Lerche and S. Theisen, Int. J. Mod. Phys. {\bf A 11} (1996) 
1929.

\bibitem{ItYa} K.~Ito and S.-K.~Yang,   Phys. Lett. {\bf B366} (1996) 165;
in \lq\lq Frontiers In Quantum Field Theory'' p. 331
(World Scientific, Singapore,1996).

\bibitem{EwFoTh}
Y.~Ohta, J. Math. Phys. {\bf 37} (1996) 6074; {\bf 38} (1997) 682; \\
H.~Ewen, K.~Foerger and S.~Theisen, 
Nucl. Phys. {\bf B485} (1997) 63; hep-th/9610049; \\
J.M. Isidro, A. Mukherjee, J.P. Nunes and  H.J. Schnitzer, hep-th/9609116; \\
M. Alishahiha, hep-th/9609157.


\bibitem{Ma}
M.~Matone,   Phys. Lett. {\bf B357} (1995) 342.

\bibitem{So}
J.~Sonnenschein, S.~Theisen and S.~Yankielowicz,
Phys. Lett. {\bf B367} (1996) 145; \\
T.~Eguchi and S.-K.~Yang, Mod. Phys. Lett. {\bf A11} (1996) 131.

\bibitem{KlLeMaVaWa}

A.~Klemm, W.~Lerche, P.~Mayer, C.~Vafa  and N.P.~Warner, 
Nucl. Phys. {\bf B477} (1996) 746.

\bibitem{LeWa}
W.~Lerche and N.P.~Warner, hep-th/9608183.

\end{thebibliography}
\end{document}